\documentclass[aps,pre,reprint,superscriptaddress,10pt]{revtex4-2}
\usepackage{amsmath}
\usepackage{amssymb}
\usepackage{graphicx}% Include figure files
\usepackage{dcolumn}% Align table columns on decimal point
\usepackage{bm}% bold math
\usepackage{hyperref}
\usepackage{courier}
\usepackage{braket}
\usepackage{float}
\usepackage[usenames,dvipsnames]{xcolor}
\hypersetup{
  colorlinks   = true,	%Colours links instead of ugly boxes
  urlcolor     = blue,	%Colour for external hyperlinks
  linkcolor    = blue,	%Colour of internal links
  citecolor   = blue	%Colour of citations#NavyBlue
}
\begin{document}
\newcommand{\ve}{\varepsilon}
%Title of paper
\title{Aging transition in coupled quantum oscillators}

\author{Biswabibek Bandyopadhyay}
\affiliation{Chaos and Complex Systems Research Laboratory, Department of Physics, University of Burdwan, Burdwan 713
  104, West Bengal, India}
\author{Tanmoy Banerjee}
\email[]{tbanerjee@phys.buruniv.ac.in}
%%\homepage[]{Your web page}
\thanks{he/his/him}
%%\altaffiliation{}
\affiliation{Chaos and Complex Systems Research Laboratory, Department of Physics, University of Burdwan, Burdwan 713
  104, West Bengal, India}

\date{\today}

\begin{abstract}
Aging transition is an emergent behavior observed in networks consisting of active (self-oscillatory) and inactive (non self-oscillatory) nodes, where the network transits from a global oscillatory state to an oscillation collapsed state when the fraction of inactive oscillators surpasses a critical value.
However, the aging transition in quantum domain has not been studied yet. In this paper we investigate the quantum manifestation of aging transition in a network of active-inactive quantum oscillators. 
We show that, unlike classical case, the quantum aging is not characterized by a complete collapse of oscillation but by sufficient reduction in the mean boson number. We identify a critical ``knee" value in the fraction of inactive oscillators around which quantum aging occurs in two different ways. Further, in stark contrast to the classical case, quantum aging transition  depends upon the nonlinear damping parameter.  
We also explain the underlying processes leading to quantum aging that have no counterpart in the classical domain.
\end{abstract}

\maketitle

%##############################################################
%###################### Introduction ##########################
%##############################################################
\section{Introduction}
\label{sec_intro}

Coupled oscillators constitute an important framework to understand several phenomena and processes in physics, biology, and engineering \cite{sync}. Depending upon the nature and strength of coupling, diverse collective behaviors are identified, such as synchronization, oscillation suppression, cluster formation, and chimera patterns \cite{sync-boca,moter,schoell_rev,scholl_CD,tanCD}. 
In this context, \citet{daido} in a seminal paper reported an interesting emergent behavior in coupled oscillators, namely the aging transition. They considered a network of globally coupled oscillators in which a fraction of oscillators are in the inactive or non-self-oscillatory state and the rest of the oscillators are in the active or self-oscillatory state. Remarkably, it was shown that, when the fraction of inactive oscillators reaches a critical value, the inactive oscillators bring down the whole network to a non-self-oscillatory steady state. The transition from macroscopic oscillatory state to a globally cessation state due to the interplay of active-inactive oscillators mediated by coupling was defined as the {\it aging transition} by \citet{daido}. Note that the aging state is different from the conventional amplitude death state: while the former state is due to the interaction between \textit{active and inactive} nodes \cite{daido,dvprep}, in the latter state \textit{active oscillators} interact among themselves to achieve oscillation cessation state \cite{aronson,kosprl,kosprep,tanpre1}.   

The aging transition is much relevant in biology and technology where maintaining oscillations is essential for the proper functioning of the system (e.g. in brain waves in neuroscience \cite{brain}, cardiopulmonary sinus rhythm of pacemaker cells \cite{cardio}, and power grids \cite{kurths_pg}) as  degradation (or aging) of some nodes in these systems may lead to irrecoverable pathological condition or system breakdown. Several studies were reported to either verify the transition in diverse systems \cite{daido-noniso,pazo-excitable,daido-pre,daido-hetero} or to propose schemes for revoking the transition \cite{kurthnat15,ryth4,sathiyaRaj}. However, all the studies are made in the classical domain (see \cite{dvprep} and references therein).             

In this work, we ask the question 
``Does the aging transition take place in the quantum domain, as well?" if yes, what will be its manifestation?
Our study is motivated by the recent endeavor of studying the well known results of classical nonlinear dynamics in the quantum domain. It is intriguing that most of the well known phenomena of classical nonlinear dynamics behaves in a counter intuitive way in the quantum domain \cite{chia}. A few prominent examples are, synchronization \cite{lee_prl,brud_prl1,brud-ann15,squeezing,blockade,brud-poch}, oscillation suppression \cite{qad1,qad2}, and symmetry-breaking states \cite{qchm,qmod,qrev,qturing,qkerr,kato}. These surprising results are attributed to the presence of inherent quantum noise, discrete energy levels, entanglement, which do not have any counterpart in classical domain. Apart from the theoretical motivation, the study of aging in the quantum domain is further motivated by the fact that system degradation or aging is inevitable in real-world quantum systems mainly due to the unwanted losses, for example, due to lossy cavity \cite{loss-cavity}, mechanical dissipation in optomechanical systems \cite{opto-revmodphys,opto-loss}, loss of trapped-ion due to collision \cite{loss-ion}, and dielectric losses of superconducting junction in circuit quantum electrodynamics \cite{cQED}. 

In this paper we show that aging transition indeed occurs in the quantum regime albeit its manifestation is different from that of the classical one. We consider a network of coupled active-inactive quantum Stuart-Landau oscillators under global diffusive coupling. Using the formalism of open quantum systems, we define the notion of active and inactive nodes in the quantum domain based on the nature of dissipators in the quantum master equation. We show that, quantum aging is not characterized by a complete cessation to a collapsed state but a rapid decrease in the mean boson number with increasing fraction of inactive oscillator. We observe that there exists a critical ``knee" value of the fraction of the inactive units around which the network exhibits two different trend of decrement in the mean boson number. We also find that the quantum aging transition is very much dependent upon the nonlinear damping parameter of individual node, which is opposite to its classical counterpart. Finally, we provide possible explanations to the quantum aging process based on the disparate relaxation process of the active and inactive oscillators.   

%#######################################################################
%#################### Classical aging transition #######################
%#######################################################################
\begin{figure}
%\vspace{0.4cm}
\includegraphics[width=.35\textwidth]{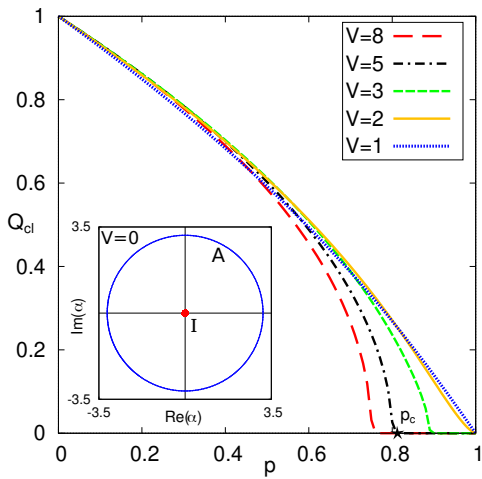}
\caption{Order parameter ($Q_{cl}$) vs. fraction of inactive oacillators ($p$) showing aging transition in the network of coupled classical oscillators. For active oscillators $G_j=a=4$ and for inactive oscillators $G_j=-b=-2$. Inset: The phase space dynamics of uncoupled ($V=0$) active (A) and inactive (I) element. Other parameters are $N=100$ and $\omega=2$.}
\label{cl_aging}
\end{figure}
\section{Classical aging transition}
For a better understanding of the quantum aging transition, let us start with a brief discussion on its classical counterpart \cite{daido}. The mathematical model of $N$ diffusively coupled identical Stuart-Landau oscillators under global coupling is given by (using a slightly different form than \cite{daido} to make it compatible with the quantum model),
\begin{equation}
\label{coupled_classical}
\dot{\alpha_j}=(-i\omega+\frac{G_j}{2}-\kappa|\alpha_j|^2)\alpha_j + \frac{V}{N}\sum_{j'=1}^N\left(\alpha_{j'}-\alpha_j\right),
\end{equation}
where $\alpha_j=x_j+iy_j$ is the complex amplitude of the $j$-th oscillator having eigenfrequency $\omega$ ($j=1,2,\cdots N$). Here $G_j$ and $\kappa$ are the linear pumping rate and nonlinear damping rate of the $j$th oscillator, respectively. $V$ is the coupling strength: for $V=0$, Eq.~\eqref{coupled_classical} represents uncoupled classical Stuart-Landau oscillators.
In the uncoupled system, $G_j$ determines the {\it activeness} of the $j$th oscillator: if $G_j>0$ the oscillator is {\it active} and shows limit cycle oscillation, on the other hand, if $G_j<0$ the oscillator is {\it inactive}, i.e., non self-oscillatory.  

The network of $N$ coupled oscillators consists of two groups of nodes: $N_a$ nodes ($N_a\le N$) are active and  $N_i$ nodes are inactive, with $N_a+N_i=N$. In the active group, an oscillator has $G_j=a>0$ ($j \in \{1, 2, ..., N_a\}$) and in the inactive group, a node has $G_j=-b<0$ ($j \in \{N_a+1, ..., N\}$).
We define the \textit{fraction of inactive nodes in the network} as $p=\frac{N_i}{N}=\frac{N-N_a}{N}$. The aging transition can be tracked using the normalized order parameter $Q_{cl}=\frac{|Z(p)|}{|Z(0)|}$, where $Z=\frac{1}{N}\sum_{j=1}^N\alpha_j$. Here $Z(p)$ and $Z(0)$ are the values of $Z$ for the fraction of inactive oscillators equal to $p$ and zero, respectively. Using a linear stability analysis, it can be shown that beyond a critical coupling strength ($V_c$), if $p$ goes beyond a critical value, given by 
\begin{equation}\label{ag}
p_c=\frac{a(2V+b)}{2(a+b)V},
\end{equation}
the network collapses into a global non self-oscillatory state \cite{daido}. From \eqref{ag} the critical coupling strength is given by $V_c=\frac{a}{2}$. The result is demonstrated in Fig.\ref{cl_aging} for $a=4$ and $b=2$. It shows that for $V>V_c(=2)$, aging transition occurs (i.e., $Q_{cl}$ becomes zero from a nonzero value) even if the fraction of inactive oscillator $p<1$. Also, for $V>V_c$ the critical value of $p$ (i.e., $p_c$) decreases with increasing $V$. Interestingly, note that \eqref{ag} does not depend upon $\kappa$, i.e., classical aging transition does not depend upon the nonlinear damping parameter.

%#######################################################################
%#################### Quantum aging transition #######################
%#######################################################################
\section{Quantum aging transition}
%==============================================================
%==============================================================
\subsection{Model}
Quantum master equation of $N$ coupled quantum Stuart-Landau oscillators under diffusive global coupling is given by \cite{qad1},
\begin{align}
\label{coupled_master}
\dot{\rho}&=\sum_{j=1}^N\left(-i[H,\rho]+\underbrace{G_j\mathcal{D}[O_j](\rho)}+\kappa\mathcal{D}[{a_j}^2](\rho)\right) \nonumber \\
&+ \frac{V}{N}\sum_{j=1}^N\sum_{j'=1}^N{}^{'} \mathcal{D}[a_j - a_{j'}](\rho), 
\end{align}
where $H=\omega {a_j}^\dag a_j$. $a_j$ and ${a_j}^{\dag}$ are bosonic annihilation and creation operators of the $j$-th oscillator, respectively. $\mathcal{D}[\hat{L}]$ is called the Lindblad dissipator having the form $\mathcal{D}[\hat{L}](\rho)=\hat{L}\rho \hat{L}^\dag-\frac{1}{2}\{\hat{L}^\dag \hat{L},\rho \}$, where $\hat{L}$ is an operator (without any loss of generality we set $\hbar=1$). The operator $O_j$ of the second term (shown in underbrace) is introduced to bring the notion of active-inactive element (discussed in the next paragraph). In the classical limit, i.e., $G_j>\kappa$, the master equation \eqref{coupled_master} is equivalent to the classical amplitude equation \eqref{coupled_classical} by the relation: $\dot{\braket{a}}=\mbox{Tr}(\dot \rho a)$. Here $\sum_{j'}{}^{'}$ indicates that the sum does not include the condition $j'=j$. 

Let us introduce the notion of {\it active} and {\it inactive} elements into the quantum master equation. We identify that this can be determined by the operator of the Lindblad dissipator associated with the coefficient $G_j$ (shown in \eqref{coupled_master} in underbrace) in the following manner
%\begin{equation}
$$O_j = 
\begin{cases}
{a_j}^\dag & \text{for active oscillators,} \\[10pt]
{a_j} & \text{for inactive oscillators.}
\end{cases}$$
%\end{equation} 
That is for $O_j={a_j}^\dag$, the dissipator $G_j\mathcal{D}[O_j](\rho)$ in Eq.~\eqref{coupled_master} describes a single boson (e.g. photon, phonon, etc.) gain with a rate of $G_j$, therefore, it gives rise to a stable quantum limit cycle for the $j$th oscillator, making it a quantum active element. However, $O_j={a_j}$ gives a single boson loss with a rate of $G_j$ and the $j$th oscillator exhibits quantum non-oscillatory or inactive behavior. 
Similar to the case of classical system, the whole network is divided into two groups --- one group consists of $N_a$ active elements and the other consists of $N_i$ inactive elements ($N_i=N-N_a$). Also, as usual $p=\frac{N_i}{N}$ denotes the fraction of inactive nodes present in the network.
Eq.\eqref{coupled_master} was studied by \citet{qad1} in the context of quantum amplitude death, where all the oscillators were taken as active (i.e., $O_j={a_j}^\dag$): it was shown that a parameter mismatch among the coupled oscillators gives rise to quantum amplitude death.

\begin{figure}
%\vspace{0.4cm}
\includegraphics[width=.4\textwidth]{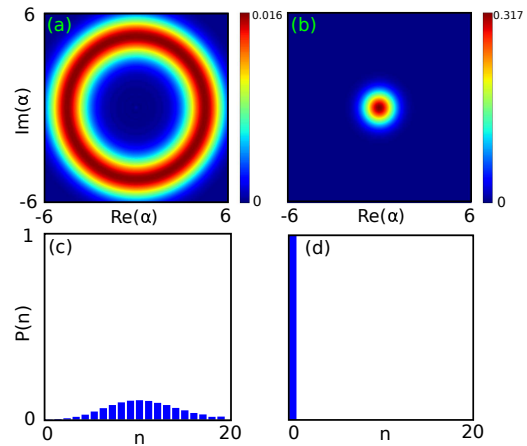}
\caption{(a,b) Wigner function distribution in phase space and (c,d) occupation probability in Fock states for uncoupled (a,c) active oscillator ($G_j=4$ and $O_j={a_j}^\dag$) and (b,d) inactive oscillator ($G_j=2$ and $O_j=a$). Other parameters are $\omega=2$ and $\kappa=0.2$.}
\label{uncoupled_quantum}
\end{figure}

For the uncoupled state ($V=0$), the Wigner-function phase space representation of {\it active} and {\it inactive} elements are shown in Fig.~\ref{uncoupled_quantum}(a) and Fig.~\ref{uncoupled_quantum}(b), respectively. The ring shaped Wigner function \cite{carmichael} of Fig.~\ref{uncoupled_quantum}(a) represents a quantum limit cycle ($G_j=4, O_j={a_j}^\dag$), i.e., an active element. A probability blob at the origin of Fig.~\ref{uncoupled_quantum}(b) (with $G_j=2, O_j={a_j}$) indicates that the element is non-self-oscillatory or inactive. The corresponding occupation probability distribution in Fock state are shown in Fig.~\ref{uncoupled_quantum} (c) (for active elements) and Fig.~\ref{uncoupled_quantum} (d) (for inactive elements). In the inactive element, indeed, ground state is the only occupied state, whereas, for the active element, higher Fock levels are more populated.

For a large number of oscillators, one can factorize the density matrix of the many-body system as $\rho \approx \otimes_{j=1}^N \rho_j$. This corresponds to the mean-field approximation and reduces the master equation Eq.~\eqref{coupled_master} to a set of master equations for each oscillator interacting with the mean-field as follows \cite{qad1}:
\begin{align}
\label{mf_master}
\dot{\rho_j}&=-i[\omega {a_j}^\dag a_j,\rho_j]+G_j\mathcal{D}[O_j](\rho_j)+\kappa\mathcal{D}[{a_j}^2](\rho_j) \nonumber \\
&+\frac{2V(N-1)}{N}\mathcal{D}[a_j](\rho)+V\left(A[{a_j}^\dag,\rho_j]-A^*[a_j,\rho_j]\right), 
\end{align}
where $A$ and $A^*$ have the forms: $A=\frac{1}{N}\sum_{j'=1}^{'N}\braket{a_{j'}}_j$ and $A^*=\frac{1}{N}\sum_{j'=1}^{'N}\braket{{a_{j'}^\dag}}_j$. Here $\braket{\hat{L}}_j$ denotes the average of operator $\hat{L}$ with respect to the one-body density matrix $\rho_j$.

%==============================================================
%==============================================================
\subsection{Results and discussion}
\begin{figure}
%\vspace{0.4cm}
%\hspace{-3.8cm}
\includegraphics[width=.49\textwidth]{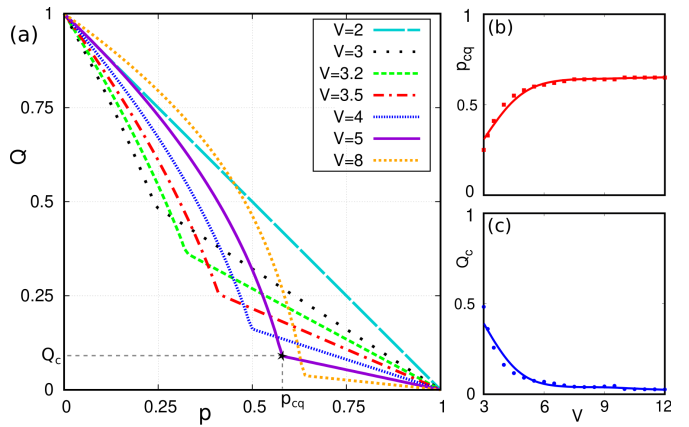}
\caption{(a) Quantum aging transition: Variation of order parameter $Q$ with $p$ for different values of coupling strength $V$. On the curve for $V=5$, the star mark denotes the knee point of the curve. Critical values of $p$ and $Q$ corresponding to the ``knee" point are denoted as $p_{cq}$ and $Q_{c}$, respectively. (b) The variation of $p_{cq}$ with coupling strength $V$. (c) Variation of $Q_{c}$ with coupling strength $V$. Other parameters are $\omega=2$ and $\kappa=0.2$.}
\label{nmf_vs_p}
\end{figure}
We numerically solve Eq.~\eqref{mf_master} by self-consistent method using QuTiP \cite{qutip1,qutip}. As in the case of classical network, here also we take  $G_j=4$ for the active elements $[j \in \{1, 2, ..., N_a\}]$, and $G_j=2$ for the inactive elements $[j \in \{N_a+1, ..., N\}]$. In the network, we distinguish oscillatory and oscillation collapsed state by computing the mean boson number per oscillator: $\bar{n}_{mf}=\frac{1}{N}\sum_j\braket{{a_j}^\dag a_j}$. Based on this, we define  an order parameter, namely the normalized mean boson number $$Q=\frac{\bar{n}_{mf}(p)}{\bar{n}_{mf}(0)},$$ 
where, $\bar{n}_{mf}(p)$ is the mean boson number per oscillator for a particular $p$ value and  $\bar{n}_{mf}(0)$ is the same for $p=0$.

To observe the variation of average mean boson number with increasing $p$ value we plot $Q$ with $p$ in Fig.~\ref{nmf_vs_p}(a) for a set of coupling strength $V$.
For $V\le 2.73$, $Q$ decreases monotonically with increasing $p$ that has a resemblance with the classical case (cf. Fig.~\ref{cl_aging}).
However, beyond a critical value of $V\approx 2.73$, the decrement rate of $Q$ shows two characteristic zones: in the first part $Q$ decreases rapidly with increasing $p$, however, beyond a critical value $p_{cq}$, the curve takes the form of an almost inclined straight line. We call the point on the curve separating the sharp fall and the inclined straight region as ``knee" point. In Fig.\ref{nmf_vs_p}(a) the knee point of the curve corresponding to $V=5$ is shown by a star mark on it. We consider these knee points as aging transition points and denote the $p$ values corresponding to these knee points as $p_{cq}$ and the corresponding order parameter as $Q_c$. With a further increment of $p$, the curve gradually approaches $Q=0$ due to the presence of more and more inactive elements in the network. We get $Q=0$ as $p$ equals unity, which is a trivial case as now all the oscillators are inactive.

\begin{figure}
%\vspace{0.4cm}
%\hspace{-3.8cm}
\includegraphics[width=.44\textwidth]{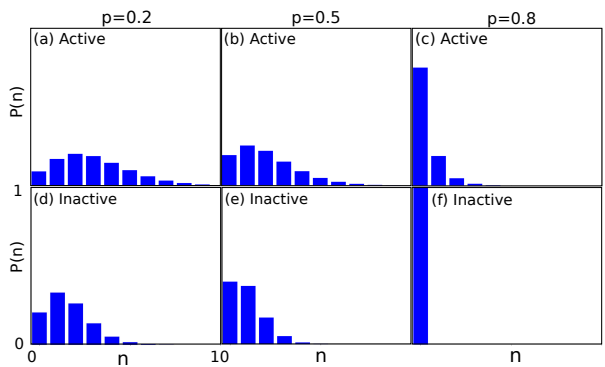}
\caption{Fock state distribution of boson numbers for active (upper row) and inactive (lower row) elements at different $p$ values (coupling strength $V=5$). (a,d) $p=0.2 (<p_{cq})$, (b,e) $p=0.5(<p_{cq})$. (c,f) $p=0.8(>p_{cq})$. Note that for $p>p_{cq}$, inactive oscillators go to the ground state, however, active oscillators still populate a few excited states. Other parameters are $\omega=2$ and $\kappa=0.2$.}
\label{fock}
\end{figure}
%Though the value of $Q$ does not fall to zero due to inherent quantum noise, the sharp decrement ensures the collective collapse which in turn confirms the occurrence of aging transition in the network of quantum oscillators.

%We can detect the knee points for a series of curves having different $V$ values. According to Fig.~\ref{nmf_vs_p}(a), each of these knee points are associated to two critical values: (i) the critical fraction of inactive elements ($p_{cl}$) and (ii) the critical order parameter ($Q_{c}$) at $p=p_{cl}$. These two values corresponding to the curve $V=5$ are shown in Fig.\ref{nmf_vs_p}(a). 
The variations of $p_{cq}$ and $Q_c$ with coupling strength $V$ are shown in Fig.~\ref{nmf_vs_p}(b) and Fig.~\ref{nmf_vs_p}(c), respectively. Fig.~\ref{nmf_vs_p}(b) shows that $p_{cq}$ at first increases with increasing $V$, then for stronger coupling strength it does not change appreciably and gets saturated. This is unlike classical case, in which $p_c$ decreases with increasing coupling strength \cite{daido}. However, $Q_c$ decreases with increasing coupling strength, which is expected as strong coupling is conducive for aging.    

The quantum aging scenario can be explained by carefully inspecting the Fock distribution. Fig.~\ref{fock}(a,d) shows the distribution of Fock states at a lower $p$ value ($p=0.2$) for active and inactive groups. Comparing this scenario with Fig.~\ref{uncoupled_quantum} (uncoupled case) one can observe that in the presence of coupling ($V>V_c)$, at first inactive elements become oscillatory and the active elements become less active. With increasing $p$, amplitude of active and inactive oscillators decreases rapidly [see Fig.~\ref{fock}(b,e) for $p=0.5$], which is manifested by the shift of probability distribution towards the ground state. Beyond the critical value $p_{cq}$, active and inactive oscillators behave in a different manner: all the inactive oscillators now completely collapsed to the ground state [see Fig.~\ref{fock}(f)], however, the active oscillators do not completely go to the ground state, rather a few lower level states near the ground state are still populated [see Fig.~\ref{fock}(e)].    
This can be understood from the nature of dissipator associated with active and inactive oscillators in Eq.~\eqref{coupled_master}. For the inactive oscillators, the dissipator (in the under brace) contains a single boson absorption term, which allows for a complete relaxation to the ground state. However, for active elements, damping is governed only by two photon absorption process, which does not allow a complete relaxation of the system to the ground state.  

\begin{figure}[t!]
%\vspace{0.4cm}
%\hspace{-3.8cm}
\includegraphics[width=.49\textwidth]{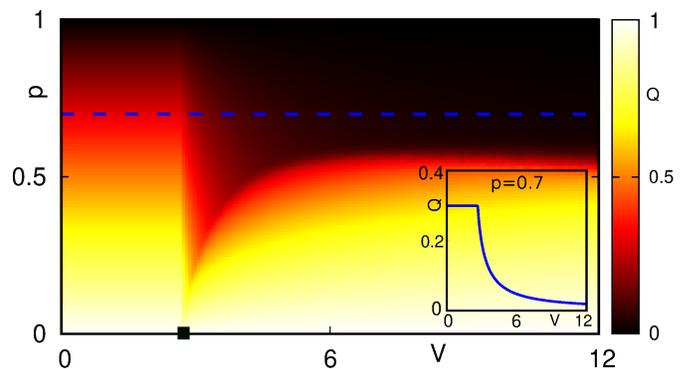}
\caption{Two parameter plot of $Q$ in $V-p$ space. Inset: The variation of $Q$ with $V$ at $p=0.7$ (along blue dashed horizontal line). The black square shows the critical value of $V$ beyond which aging can take place. Other parameters are $\omega=2$ and $\kappa=0.2$.}
\label{onepar_twopar}
\end{figure}

Next, we present the quantum aging scenario in the $V-p$ space by visualizing the variation of the normalized order parameter $Q$ (shown in Fig.~\ref{onepar_twopar}). The plot shows that up to a certain value of coupling strength $V$ (shown by a square mark on the $V$-axis), no aging transition occurs; beyond this value of $V$ the aging transition occurs with increasing $p$. Choosing $p=0.7$, we plot $Q-V$ (along the horizontal line of Fig.~\ref{onepar_twopar}), which is shown in the inset of Fig.~\ref{onepar_twopar}. It shows that beyond a critical value of $V$, the order parameter $Q$ decreases with $V$.  

\begin{figure}
%\vspace{0.4cm}
%\hspace{-3.8cm}
\includegraphics[width=.4\textwidth]{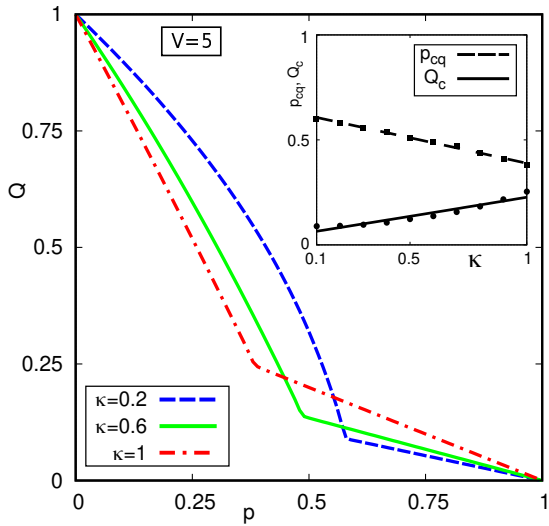}
\caption{Plot of $p-Q$ showing the effect of nonlinear damping parameter $\kappa$ on the quantum aging ($V=5$). Inset: The variation of $p_{cq}$ and $Q$ with $\kappa$.}
\label{qk}
\end{figure}
In the classical case, aging transition does not depend upon the nonlinear damping parameter $\kappa$. However, in stark contrast, we find that in the quantum case, $\kappa$ very much controls the quantum aging transition. Fig.~\ref{qk} demonstrates the variation of $Q$ with $p$ for three different exemplary values of $\kappa$ ($V=5$): it shows that a large $\kappa$ induces quantum aging at a lower $p$ value. The variation of $p_{cq}$ with $\kappa$ in the inset supports this fact. Also note that larger $\kappa$ results in a greater $Q_c$ value, i.e., nonlinearity degrades the quality of aging in the quantum aging transition. This can be attributed to the fact that, higher damping makes the oscillators to populate only a few excited state near the ground state even in the absence of aging, therefore, aging has a lesser role to bring them further closer to the ground state.

%##############################################################
%###################### Conclusions ###########################
%##############################################################
\section{Conclusions}
\label{sec_conc}
In this paper we have studied the quantum mechanical analog of aging transition. We have considered a network of active and inactive quantum oscillators and using a mean-boson number based order parameter we have  demonstrated the quantum version of the aging transition. The following are the key observations that make quantum aging transition different from its classical counterpart.
First, unlike classical case, the quantum aging transition is not manifested by the complete collapse of the network but by rapid decrease in the average mean boson number. 
%The order parameter becomes zero only when all the elements become inactive, which is a trivial case. 
Second, the quantum aging has two distinct processes associated with it: up to a critical ``knee" point, the order parameter decreases rapidly, however, beyond that the rate of decrement slows down. In the second part all the inactive elements populate the ground state, which is allowed by the single boson loss process. However, here active oscillators never completely populate the ground state, because for them relaxation is governed by the two-boson absorption process only. This scenario has no classical counterpart and makes the quantum aging process unique.
Third, unlike classical aging transition, quantum aging depends upon the nonlinear damping process. We have shown that a strong nonlinear damping makes the aging process weaker. We attribute this to the population of a few excited states in the presence of strong nonlinear damping that does not change appreciably by aging.        

The study can be implemented in experiments by realizing one boson loss (for inactive elements), one boson gain (for active elements), and two boson absorption process (for both active and inactive nodes). With the advancement of current quantum technology, we believe that trapped-ion set ups and optomechanical systems can provide a feasible experimental platform for verifying the results. Our study has revealed that the presence of a few inactive (or {\it aged}) elements in a quantum network may lead to system dysfunction or collapse. Therefore, sufficient measures have to be taken to avoid node degradation due to possible losses \cite{low-loss,low-loss2}.

\begin{acknowledgments}
B.B. acknowledges the financial assistance from the University Grants Commission (UGC), India in the form of Senior Research Fellowship (SRF). T. B. acknowledges the financial support from the Science and Engineering Research Board (SERB), Government of India, in the form of a Core Research Grant [CRG/2019/002632].
\end{acknowledgments}

%\bibliography{qaging}

\begin{thebibliography}{47}%
\makeatletter
\providecommand \@ifxundefined [1]{%
 \@ifx{#1\undefined}
}%
\providecommand \@ifnum [1]{%
 \ifnum #1\expandafter \@firstoftwo
 \else \expandafter \@secondoftwo
 \fi
}%
\providecommand \@ifx [1]{%
 \ifx #1\expandafter \@firstoftwo
 \else \expandafter \@secondoftwo
 \fi
}%
\providecommand \natexlab [1]{#1}%
\providecommand \enquote  [1]{``#1''}%
\providecommand \bibnamefont  [1]{#1}%
\providecommand \bibfnamefont [1]{#1}%
\providecommand \citenamefont [1]{#1}%
\providecommand \href@noop [0]{\@secondoftwo}%
\providecommand \href [0]{\begingroup \@sanitize@url \@href}%
\providecommand \@href[1]{\@@startlink{#1}\@@href}%
\providecommand \@@href[1]{\endgroup#1\@@endlink}%
\providecommand \@sanitize@url [0]{\catcode `\\12\catcode `\$12\catcode
  `\&12\catcode `\#12\catcode `\^12\catcode `\_12\catcode `\%12\relax}%
\providecommand \@@startlink[1]{}%
\providecommand \@@endlink[0]{}%
\providecommand \url  [0]{\begingroup\@sanitize@url \@url }%
\providecommand \@url [1]{\endgroup\@href {#1}{\urlprefix }}%
\providecommand \urlprefix  [0]{URL }%
\providecommand \Eprint [0]{\href }%
\providecommand \doibase [0]{https://doi.org/}%
\providecommand \selectlanguage [0]{\@gobble}%
\providecommand \bibinfo  [0]{\@secondoftwo}%
\providecommand \bibfield  [0]{\@secondoftwo}%
\providecommand \translation [1]{[#1]}%
\providecommand \BibitemOpen [0]{}%
\providecommand \bibitemStop [0]{}%
\providecommand \bibitemNoStop [0]{.\EOS\space}%
\providecommand \EOS [0]{\spacefactor3000\relax}%
\providecommand \BibitemShut  [1]{\csname bibitem#1\endcsname}%
\let\auto@bib@innerbib\@empty
%</preamble>
\bibitem [{\citenamefont {Strogatz}(2012)}]{sync}%
  \BibitemOpen
  \bibfield  {author} {\bibinfo {author} {\bibfnamefont {S.~H.}\ \bibnamefont
  {Strogatz}},\ }\href@noop {} {\emph {\bibinfo {title} {Sync: How Order
  Emerges from Chaos In the Universe, Nature, and Daily Life}}}\ (\bibinfo
  {publisher} {Hyperion, New York},\ \bibinfo {year} {2012})\BibitemShut
  {NoStop}%
\bibitem [{\citenamefont {Boccaletti}\ \emph {et~al.}(2018)\citenamefont
  {Boccaletti}, \citenamefont {Pisarchik}, \citenamefont {del Genio},\ and\
  \citenamefont {Amann}}]{sync-boca}%
  \BibitemOpen
  \bibfield  {author} {\bibinfo {author} {\bibfnamefont {S.}~\bibnamefont
  {Boccaletti}}, \bibinfo {author} {\bibfnamefont {A.~N.}\ \bibnamefont
  {Pisarchik}}, \bibinfo {author} {\bibfnamefont {C.~I.}\ \bibnamefont {del
  Genio}},\ and\ \bibinfo {author} {\bibfnamefont {A.}~\bibnamefont {Amann}},\
  }\href@noop {} {\emph {\bibinfo {title} {Synchronization: From Coupled
  Systems to Complex Networks}}}\ (\bibinfo  {publisher} {Cambridge University
  Press, UK},\ \bibinfo {year} {2018})\BibitemShut {NoStop}%
\bibitem [{\citenamefont {Motter}(2010)}]{moter}%
  \BibitemOpen
  \bibfield  {author} {\bibinfo {author} {\bibfnamefont {A.~E.}\ \bibnamefont
  {Motter}},\ }\bibfield  {title} {\bibinfo {title} {Spontaneous synchrony
  breaking},\ }\href@noop {} {\bibfield  {journal} {\bibinfo  {journal} {Nat.
  Phys.}\ }\textbf {\bibinfo {volume} {6}},\ \bibinfo {pages} {164} (\bibinfo
  {year} {2010})}\BibitemShut {NoStop}%
\bibitem [{\citenamefont {Sch{\"o}ll}(2016)}]{schoell_rev}%
  \BibitemOpen
  \bibfield  {author} {\bibinfo {author} {\bibfnamefont {E.}~\bibnamefont
  {Sch{\"o}ll}},\ }\bibfield  {title} {\bibinfo {title} {Synchronization
  patterns and chimera states in complex networks: Interplay of topology and
  dynamics},\ }\href@noop {} {\bibfield  {journal} {\bibinfo  {journal} {Eur.
  Phys. J. Special Topics}\ }\textbf {\bibinfo {volume} {225}},\ \bibinfo
  {pages} {891} (\bibinfo {year} {2016})}\BibitemShut {NoStop}%
\bibitem [{\citenamefont {Zakharova}\ \emph {et~al.}(2014)\citenamefont
  {Zakharova}, \citenamefont {Kapeller},\ and\ \citenamefont
  {Sch{\"{o}}ll}}]{scholl_CD}%
  \BibitemOpen
  \bibfield  {author} {\bibinfo {author} {\bibfnamefont {A.}~\bibnamefont
  {Zakharova}}, \bibinfo {author} {\bibfnamefont {M.}~\bibnamefont
  {Kapeller}},\ and\ \bibinfo {author} {\bibfnamefont {E.}~\bibnamefont
  {Sch{\"{o}}ll}},\ }\bibfield  {title} {\bibinfo {title} {Chimera death:
  Symmetry breaking in dynamical networks},\ }\href@noop {} {\bibfield
  {journal} {\bibinfo  {journal} {Phy. Rev. Lett}\ }\textbf {\bibinfo {volume}
  {112}},\ \bibinfo {pages} {154101} (\bibinfo {year} {2014})}\BibitemShut
  {NoStop}%
\bibitem [{\citenamefont {Banerjee}(2015)}]{tanCD}%
  \BibitemOpen
  \bibfield  {author} {\bibinfo {author} {\bibfnamefont {T.}~\bibnamefont
  {Banerjee}},\ }\bibfield  {title} {\bibinfo {title}
  {Mean-field-diffusion--induced death state},\ }\href@noop {} {\bibfield
  {journal} {\bibinfo  {journal} {EPL}\ }\textbf {\bibinfo {volume} {110}},\
  \bibinfo {pages} {60003} (\bibinfo {year} {2015})}\BibitemShut {NoStop}%
\bibitem [{\citenamefont {Daido}\ and\ \citenamefont
  {Nakanishi}(2004)}]{daido}%
  \BibitemOpen
  \bibfield  {author} {\bibinfo {author} {\bibfnamefont {H.}~\bibnamefont
  {Daido}}\ and\ \bibinfo {author} {\bibfnamefont {K.}~\bibnamefont
  {Nakanishi}},\ }\bibfield  {title} {\bibinfo {title} {Aging transition and
  universal scaling in oscillator networks},\ }\href@noop {} {\bibfield
  {journal} {\bibinfo  {journal} {Phys. Rev. Lett.}\ }\textbf {\bibinfo
  {volume} {93}},\ \bibinfo {pages} {104101} (\bibinfo {year}
  {2004})}\BibitemShut {NoStop}%
\bibitem [{\citenamefont {Zou}\ \emph {et~al.}(2021)\citenamefont {Zou},
  \citenamefont {Senthilkumar}, \citenamefont {Zhan},\ and\ \citenamefont
  {Kurths}}]{dvprep}%
  \BibitemOpen
  \bibfield  {author} {\bibinfo {author} {\bibfnamefont {W.}~\bibnamefont
  {Zou}}, \bibinfo {author} {\bibfnamefont {D.~V.}\ \bibnamefont
  {Senthilkumar}}, \bibinfo {author} {\bibfnamefont {M.}~\bibnamefont {Zhan}},\
  and\ \bibinfo {author} {\bibfnamefont {J.}~\bibnamefont {Kurths}},\
  }\bibfield  {title} {\bibinfo {title} {Quenching, aging, and reviving in
  coupled dynamical networks},\ }\href@noop {} {\bibfield  {journal} {\bibinfo
  {journal} {Physics Reports}\ }\textbf {\bibinfo {volume} {931}},\ \bibinfo
  {pages} {1} (\bibinfo {year} {2021})}\BibitemShut {NoStop}%
\bibitem [{\citenamefont {Aronson}\ \emph {et~al.}(1990)\citenamefont
  {Aronson}, \citenamefont {Ermentrout},\ and\ \citenamefont
  {Kopell}}]{aronson}%
  \BibitemOpen
  \bibfield  {author} {\bibinfo {author} {\bibfnamefont {D.~G.}\ \bibnamefont
  {Aronson}}, \bibinfo {author} {\bibfnamefont {G.~B.}\ \bibnamefont
  {Ermentrout}},\ and\ \bibinfo {author} {\bibfnamefont {N.}~\bibnamefont
  {Kopell}},\ }\bibfield  {title} {\bibinfo {title} {Amplitude response of
  coupled oscillators},\ }\href@noop {} {\bibfield  {journal} {\bibinfo
  {journal} {Physica D}\ }\textbf {\bibinfo {volume} {41}},\ \bibinfo {pages}
  {403} (\bibinfo {year} {1990})}\BibitemShut {NoStop}%
\bibitem [{\citenamefont {Koseska}\ \emph
  {et~al.}(2013{\natexlab{a}})\citenamefont {Koseska}, \citenamefont {Volkov},\
  and\ \citenamefont {Kurths}}]{kosprl}%
  \BibitemOpen
  \bibfield  {author} {\bibinfo {author} {\bibfnamefont {A.}~\bibnamefont
  {Koseska}}, \bibinfo {author} {\bibfnamefont {E.}~\bibnamefont {Volkov}},\
  and\ \bibinfo {author} {\bibfnamefont {J.}~\bibnamefont {Kurths}},\
  }\bibfield  {title} {\bibinfo {title} {Transition from amplitude to
  oscillation death via turing bifurcation},\ }\href@noop {} {\bibfield
  {journal} {\bibinfo  {journal} {Phys. Rev. Lett}\ }\textbf {\bibinfo {volume}
  {111}},\ \bibinfo {pages} {024103} (\bibinfo {year}
  {2013}{\natexlab{a}})}\BibitemShut {NoStop}%
\bibitem [{\citenamefont {Koseska}\ \emph
  {et~al.}(2013{\natexlab{b}})\citenamefont {Koseska}, \citenamefont {Volkov},\
  and\ \citenamefont {Kurths}}]{kosprep}%
  \BibitemOpen
  \bibfield  {author} {\bibinfo {author} {\bibfnamefont {A.}~\bibnamefont
  {Koseska}}, \bibinfo {author} {\bibfnamefont {E.}~\bibnamefont {Volkov}},\
  and\ \bibinfo {author} {\bibfnamefont {J.}~\bibnamefont {Kurths}},\
  }\bibfield  {title} {\bibinfo {title} {Oscillation quenching mechanisms:
  Amplitude vs oscillation death},\ }\href@noop {} {\bibfield  {journal}
  {\bibinfo  {journal} {Phys. Reports}\ }\textbf {\bibinfo {volume} {531}},\
  \bibinfo {pages} {173} (\bibinfo {year} {2013}{\natexlab{b}})}\BibitemShut
  {NoStop}%
\bibitem [{\citenamefont {Banerjee}\ and\ \citenamefont
  {Ghosh}(2014)}]{tanpre1}%
  \BibitemOpen
  \bibfield  {author} {\bibinfo {author} {\bibfnamefont {T.}~\bibnamefont
  {Banerjee}}\ and\ \bibinfo {author} {\bibfnamefont {D.}~\bibnamefont
  {Ghosh}},\ }\bibfield  {title} {\bibinfo {title} {Transition from amplitude
  to oscillation death under mean-field diffusive coupling},\ }\href@noop {}
  {\bibfield  {journal} {\bibinfo  {journal} {Phys. Rev. E}\ }\textbf {\bibinfo
  {volume} {89}},\ \bibinfo {pages} {052912} (\bibinfo {year}
  {2014})}\BibitemShut {NoStop}%
\bibitem [{\citenamefont {Lisman}\ and\ \citenamefont {Buzsaki}(2008)}]{brain}%
  \BibitemOpen
  \bibfield  {author} {\bibinfo {author} {\bibfnamefont {J.}~\bibnamefont
  {Lisman}}\ and\ \bibinfo {author} {\bibfnamefont {G.}~\bibnamefont
  {Buzsaki}},\ }\bibfield  {title} {\bibinfo {title} {A neural coding scheme
  formed by the combined function of gamma and theta oscillations.},\
  }\href@noop {} {\bibfield  {journal} {\bibinfo  {journal} {Schizophr. Bull.}\
  }\textbf {\bibinfo {volume} {34}},\ \bibinfo {pages} {974} (\bibinfo {year}
  {2008})}\BibitemShut {NoStop}%
\bibitem [{\citenamefont {Jalife}\ \emph {et~al.}(1998)\citenamefont {Jalife},
  \citenamefont {Gray}, \citenamefont {Morley},\ and\ \citenamefont
  {Davidenko}}]{cardio}%
  \BibitemOpen
  \bibfield  {author} {\bibinfo {author} {\bibfnamefont {J.}~\bibnamefont
  {Jalife}}, \bibinfo {author} {\bibfnamefont {R.~A.}\ \bibnamefont {Gray}},
  \bibinfo {author} {\bibfnamefont {G.~E.}\ \bibnamefont {Morley}},\ and\
  \bibinfo {author} {\bibfnamefont {J.~M.}\ \bibnamefont {Davidenko}},\
  }\bibfield  {title} {\bibinfo {title} {Self-organization and the dynamical
  nature of ventricular fibrillation},\ }\href@noop {} {\bibfield  {journal}
  {\bibinfo  {journal} {Chaos}\ }\textbf {\bibinfo {volume} {8}},\ \bibinfo
  {pages} {79} (\bibinfo {year} {1998})}\BibitemShut {NoStop}%
\bibitem [{\citenamefont {Menck}\ \emph {et~al.}(2014)\citenamefont {Menck},
  \citenamefont {Heitzig}, \citenamefont {Kurths},\ and\ \citenamefont
  {Schellnhuber}}]{kurths_pg}%
  \BibitemOpen
  \bibfield  {author} {\bibinfo {author} {\bibfnamefont {P.~J.}\ \bibnamefont
  {Menck}}, \bibinfo {author} {\bibfnamefont {J.}~\bibnamefont {Heitzig}},
  \bibinfo {author} {\bibfnamefont {J.}~\bibnamefont {Kurths}},\ and\ \bibinfo
  {author} {\bibfnamefont {H.~J.}\ \bibnamefont {Schellnhuber}},\ }\bibfield
  {title} {\bibinfo {title} {How dead ends undermine power grid stability},\
  }\href@noop {} {\bibfield  {journal} {\bibinfo  {journal} {Nat. Commun.}\
  }\textbf {\bibinfo {volume} {5}},\ \bibinfo {pages} {3969} (\bibinfo {year}
  {2014})}\BibitemShut {NoStop}%
\bibitem [{\citenamefont {Daido}\ and\ \citenamefont
  {Nakanishi}(2006)}]{daido-noniso}%
  \BibitemOpen
  \bibfield  {author} {\bibinfo {author} {\bibfnamefont {H.}~\bibnamefont
  {Daido}}\ and\ \bibinfo {author} {\bibfnamefont {K.}~\bibnamefont
  {Nakanishi}},\ }\bibfield  {title} {\bibinfo {title} {Diffusion-induced
  inhomogeneity in globally coupled oscillators: Swing-by mechanism},\
  }\href@noop {} {\bibfield  {journal} {\bibinfo  {journal} {Phys. Rev. Lett.}\
  }\textbf {\bibinfo {volume} {96}},\ \bibinfo {pages} {054101} (\bibinfo
  {year} {2006})}\BibitemShut {NoStop}%
\bibitem [{\citenamefont {Paz\'o}\ and\ \citenamefont
  {Montbri\'o}(2006)}]{pazo-excitable}%
  \BibitemOpen
  \bibfield  {author} {\bibinfo {author} {\bibfnamefont {D.}~\bibnamefont
  {Paz\'o}}\ and\ \bibinfo {author} {\bibfnamefont {E.}~\bibnamefont
  {Montbri\'o}},\ }\bibfield  {title} {\bibinfo {title} {Universal behavior in
  populations composed of excitable and self-oscillatory elements},\
  }\href@noop {} {\bibfield  {journal} {\bibinfo  {journal} {Phys. Rev. E}\
  }\textbf {\bibinfo {volume} {73}},\ \bibinfo {pages} {055202} (\bibinfo
  {year} {2006})}\BibitemShut {NoStop}%
\bibitem [{\citenamefont {Daido}\ and\ \citenamefont
  {Nakanishi}(2007)}]{daido-pre}%
  \BibitemOpen
  \bibfield  {author} {\bibinfo {author} {\bibfnamefont {H.}~\bibnamefont
  {Daido}}\ and\ \bibinfo {author} {\bibfnamefont {K.}~\bibnamefont
  {Nakanishi}},\ }\bibfield  {title} {\bibinfo {title} {Aging and clustering in
  globally coupled oscillators},\ }\href@noop {} {\bibfield  {journal}
  {\bibinfo  {journal} {Phys. Rev. E}\ }\textbf {\bibinfo {volume} {75}},\
  \bibinfo {pages} {056206} (\bibinfo {year} {2007})}\BibitemShut {NoStop}%
\bibitem [{\citenamefont {Tanaka}\ \emph {et~al.}(2014)\citenamefont {Tanaka},
  \citenamefont {Morino}, \citenamefont {Daido},\ and\ \citenamefont
  {Aihara}}]{daido-hetero}%
  \BibitemOpen
  \bibfield  {author} {\bibinfo {author} {\bibfnamefont {G.}~\bibnamefont
  {Tanaka}}, \bibinfo {author} {\bibfnamefont {K.}~\bibnamefont {Morino}},
  \bibinfo {author} {\bibfnamefont {H.}~\bibnamefont {Daido}},\ and\ \bibinfo
  {author} {\bibfnamefont {K.}~\bibnamefont {Aihara}},\ }\bibfield  {title}
  {\bibinfo {title} {Dynamical robustness of coupled heterogeneous
  oscillators},\ }\href@noop {} {\bibfield  {journal} {\bibinfo  {journal}
  {Phys. Rev. E}\ }\textbf {\bibinfo {volume} {89}},\ \bibinfo {pages} {052906}
  (\bibinfo {year} {2014})}\BibitemShut {NoStop}%
\bibitem [{\citenamefont {Zou}\ \emph {et~al.}(2015)\citenamefont {Zou},
  \citenamefont {Senthilkumar}, \citenamefont {Nagao}, \citenamefont {Kiss},
  \citenamefont {Tang}, \citenamefont {Koseska}, \citenamefont {Duan},\ and\
  \citenamefont {Kurths}}]{kurthnat15}%
  \BibitemOpen
  \bibfield  {author} {\bibinfo {author} {\bibfnamefont {W.}~\bibnamefont
  {Zou}}, \bibinfo {author} {\bibfnamefont {D.~V.}\ \bibnamefont
  {Senthilkumar}}, \bibinfo {author} {\bibfnamefont {R.}~\bibnamefont {Nagao}},
  \bibinfo {author} {\bibfnamefont {I.~Z.}\ \bibnamefont {Kiss}}, \bibinfo
  {author} {\bibfnamefont {Y.}~\bibnamefont {Tang}}, \bibinfo {author}
  {\bibfnamefont {A.}~\bibnamefont {Koseska}}, \bibinfo {author} {\bibfnamefont
  {J.}~\bibnamefont {Duan}},\ and\ \bibinfo {author} {\bibfnamefont
  {J.}~\bibnamefont {Kurths}},\ }\bibfield  {title} {\bibinfo {title}
  {Restoration of rhythmicity in diffusively coupled dynamical networks},\
  }\href@noop {} {\bibfield  {journal} {\bibinfo  {journal} {Nat. Commun.}\
  }\textbf {\bibinfo {volume} {6}},\ \bibinfo {pages} {7709} (\bibinfo {year}
  {2015})}\BibitemShut {NoStop}%
\bibitem [{\citenamefont {Morino}\ \emph {et~al.}(2013)\citenamefont {Morino},
  \citenamefont {Tanaka},\ and\ \citenamefont {Aihara}}]{ryth4}%
  \BibitemOpen
  \bibfield  {author} {\bibinfo {author} {\bibfnamefont {K.}~\bibnamefont
  {Morino}}, \bibinfo {author} {\bibfnamefont {G.}~\bibnamefont {Tanaka}},\
  and\ \bibinfo {author} {\bibfnamefont {K.}~\bibnamefont {Aihara}},\
  }\bibfield  {title} {\bibinfo {title} {Efficient recovery of dynamic behavior
  in coupled oscillator networks},\ }\href@noop {} {\bibfield  {journal}
  {\bibinfo  {journal} {Phys. Rev. E}\ }\textbf {\bibinfo {volume} {88}},\
  \bibinfo {pages} {032909} (\bibinfo {year} {2013})}\BibitemShut {NoStop}%
\bibitem [{\citenamefont {Sathiyadevi}\ \emph {et~al.}(2022)\citenamefont
  {Sathiyadevi}, \citenamefont {Premraj}, \citenamefont {Banerjee},
  \citenamefont {Zheng},\ and\ \citenamefont {Lakshmanan}}]{sathiyaRaj}%
  \BibitemOpen
  \bibfield  {author} {\bibinfo {author} {\bibfnamefont {K.}~\bibnamefont
  {Sathiyadevi}}, \bibinfo {author} {\bibfnamefont {D.}~\bibnamefont
  {Premraj}}, \bibinfo {author} {\bibfnamefont {T.}~\bibnamefont {Banerjee}},
  \bibinfo {author} {\bibfnamefont {Z.}~\bibnamefont {Zheng}},\ and\ \bibinfo
  {author} {\bibfnamefont {M.}~\bibnamefont {Lakshmanan}},\ }\bibfield  {title}
  {\bibinfo {title} {Aging transition under discrete time-dependent coupling:
  Restoring rhythmicity from aging},\ }\href@noop {} {\bibfield  {journal}
  {\bibinfo  {journal} {Chaos, Solitons {\&} Fractals}\ }\textbf {\bibinfo
  {volume} {157}},\ \bibinfo {pages} {111944} (\bibinfo {year}
  {2022})}\BibitemShut {NoStop}%
\bibitem [{\citenamefont {Chia}\ \emph {et~al.}(2020)\citenamefont {Chia},
  \citenamefont {Kwek},\ and\ \citenamefont {Noh}}]{chia}%
  \BibitemOpen
  \bibfield  {author} {\bibinfo {author} {\bibfnamefont {A.}~\bibnamefont
  {Chia}}, \bibinfo {author} {\bibfnamefont {L.~C.}\ \bibnamefont {Kwek}},\
  and\ \bibinfo {author} {\bibfnamefont {C.}~\bibnamefont {Noh}},\ }\bibfield
  {title} {\bibinfo {title} {Relaxation oscillations and frequency entrainment
  in quantum mechanics},\ }\href@noop {} {\bibfield  {journal} {\bibinfo
  {journal} {Phys. Rev. E}\ }\textbf {\bibinfo {volume} {102}},\ \bibinfo
  {pages} {042213} (\bibinfo {year} {2020})}\BibitemShut {NoStop}%
\bibitem [{\citenamefont {Lee}\ and\ \citenamefont
  {Sadeghpour}(2013)}]{lee_prl}%
  \BibitemOpen
  \bibfield  {author} {\bibinfo {author} {\bibfnamefont {T.~E.}\ \bibnamefont
  {Lee}}\ and\ \bibinfo {author} {\bibfnamefont {H.~R.}\ \bibnamefont
  {Sadeghpour}},\ }\bibfield  {title} {\bibinfo {title} {Quantum
  synchronization of quantum van der pol oscillators with trapped ions},\
  }\href@noop {} {\bibfield  {journal} {\bibinfo  {journal} {Phys. Rev. Lett.}\
  }\textbf {\bibinfo {volume} {111}},\ \bibinfo {pages} {234101} (\bibinfo
  {year} {2013})}\BibitemShut {NoStop}%
\bibitem [{\citenamefont {Walter}\ \emph {et~al.}(2014)\citenamefont {Walter},
  \citenamefont {Nunnenkamp},\ and\ \citenamefont {Bruder}}]{brud_prl1}%
  \BibitemOpen
  \bibfield  {author} {\bibinfo {author} {\bibfnamefont {S.}~\bibnamefont
  {Walter}}, \bibinfo {author} {\bibfnamefont {A.}~\bibnamefont {Nunnenkamp}},\
  and\ \bibinfo {author} {\bibfnamefont {C.}~\bibnamefont {Bruder}},\
  }\bibfield  {title} {\bibinfo {title} {Quantum synchronization of a driven
  self-sustained oscillator},\ }\href@noop {} {\bibfield  {journal} {\bibinfo
  {journal} {Phys. Rev. Lett.}\ }\textbf {\bibinfo {volume} {112}},\ \bibinfo
  {pages} {094102} (\bibinfo {year} {2014})}\BibitemShut {NoStop}%
\bibitem [{\citenamefont {Walter}\ \emph {et~al.}(2015)\citenamefont {Walter},
  \citenamefont {Nunnenkamp},\ and\ \citenamefont {Bruder}}]{brud-ann15}%
  \BibitemOpen
  \bibfield  {author} {\bibinfo {author} {\bibfnamefont {S.}~\bibnamefont
  {Walter}}, \bibinfo {author} {\bibfnamefont {A.}~\bibnamefont {Nunnenkamp}},\
  and\ \bibinfo {author} {\bibfnamefont {C.}~\bibnamefont {Bruder}},\
  }\bibfield  {title} {\bibinfo {title} {Quantum synchronization of two van der
  pol oscillators},\ }\href@noop {} {\bibfield  {journal} {\bibinfo  {journal}
  {Ann. der. Phys.}\ }\textbf {\bibinfo {volume} {527}},\ \bibinfo {pages}
  {131} (\bibinfo {year} {2015})}\BibitemShut {NoStop}%
\bibitem [{\citenamefont {Sonar}\ \emph {et~al.}(2018)\citenamefont {Sonar},
  \citenamefont {Hajdu{\v{s}}ek}, \citenamefont {Mukherjee}, \citenamefont
  {Fazio}, \citenamefont {Vedral}, \citenamefont {Vinjanampathy},\ and\
  \citenamefont {Kwek}}]{squeezing}%
  \BibitemOpen
  \bibfield  {author} {\bibinfo {author} {\bibfnamefont {S.}~\bibnamefont
  {Sonar}}, \bibinfo {author} {\bibfnamefont {M.}~\bibnamefont
  {Hajdu{\v{s}}ek}}, \bibinfo {author} {\bibfnamefont {M.}~\bibnamefont
  {Mukherjee}}, \bibinfo {author} {\bibfnamefont {R.}~\bibnamefont {Fazio}},
  \bibinfo {author} {\bibfnamefont {V.}~\bibnamefont {Vedral}}, \bibinfo
  {author} {\bibfnamefont {S.}~\bibnamefont {Vinjanampathy}},\ and\ \bibinfo
  {author} {\bibfnamefont {L.}~\bibnamefont {Kwek}},\ }\bibfield  {title}
  {\bibinfo {title} {Squeezing enhances quantum synchronization},\ }\href@noop
  {} {\bibfield  {journal} {\bibinfo  {journal} {Phys. Rev. Lett.}\ }\textbf
  {\bibinfo {volume} {120}},\ \bibinfo {pages} {163601} (\bibinfo {year}
  {2018})}\BibitemShut {NoStop}%
\bibitem [{\citenamefont {L{\"{o}}rch}\ \emph {et~al.}(2017)\citenamefont
  {L{\"{o}}rch}, \citenamefont {Nigg}, \citenamefont {Nunnenkamp},
  \citenamefont {Tiwari},\ and\ \citenamefont {Bruder}}]{blockade}%
  \BibitemOpen
  \bibfield  {author} {\bibinfo {author} {\bibfnamefont {N.}~\bibnamefont
  {L{\"{o}}rch}}, \bibinfo {author} {\bibfnamefont {S.~E.}\ \bibnamefont
  {Nigg}}, \bibinfo {author} {\bibfnamefont {A.}~\bibnamefont {Nunnenkamp}},
  \bibinfo {author} {\bibfnamefont {R.~P.}\ \bibnamefont {Tiwari}},\ and\
  \bibinfo {author} {\bibfnamefont {C.}~\bibnamefont {Bruder}},\ }\bibfield
  {title} {\bibinfo {title} {Quantum synchronization blockade: Energy
  quantization hinders synchronization of identical oscillators},\ }\href@noop
  {} {\bibfield  {journal} {\bibinfo  {journal} {Phys. Rev. Lett}\ }\textbf
  {\bibinfo {volume} {118}},\ \bibinfo {pages} {243602} (\bibinfo {year}
  {2017})}\BibitemShut {NoStop}%
\bibitem [{\citenamefont {L\"orch}\ \emph {et~al.}(2016)\citenamefont
  {L\"orch}, \citenamefont {Amitai}, \citenamefont {Nunnenkamp},\ and\
  \citenamefont {Bruder}}]{brud-poch}%
  \BibitemOpen
  \bibfield  {author} {\bibinfo {author} {\bibfnamefont {N.}~\bibnamefont
  {L\"orch}}, \bibinfo {author} {\bibfnamefont {E.}~\bibnamefont {Amitai}},
  \bibinfo {author} {\bibfnamefont {A.}~\bibnamefont {Nunnenkamp}},\ and\
  \bibinfo {author} {\bibfnamefont {C.}~\bibnamefont {Bruder}},\ }\bibfield
  {title} {\bibinfo {title} {Genuine quantum signatures in synchronization of
  anharmonic self-oscillators},\ }\href@noop {} {\bibfield  {journal} {\bibinfo
   {journal} {Phys. Rev. Lett.}\ }\textbf {\bibinfo {volume} {117}},\ \bibinfo
  {pages} {073601} (\bibinfo {year} {2016})}\BibitemShut {NoStop}%
\bibitem [{\citenamefont {Ishibashi}\ and\ \citenamefont
  {Kanamoto}(2017)}]{qad1}%
  \BibitemOpen
  \bibfield  {author} {\bibinfo {author} {\bibfnamefont {K.}~\bibnamefont
  {Ishibashi}}\ and\ \bibinfo {author} {\bibfnamefont {R.}~\bibnamefont
  {Kanamoto}},\ }\bibfield  {title} {\bibinfo {title} {Oscillation collapse in
  coupled quantum van der pol oscillators},\ }\href@noop {} {\bibfield
  {journal} {\bibinfo  {journal} {Phys. Rev. E}\ }\textbf {\bibinfo {volume}
  {96}},\ \bibinfo {pages} {052210} (\bibinfo {year} {2017})}\BibitemShut
  {NoStop}%
\bibitem [{\citenamefont {Amitai}\ \emph {et~al.}(2018)\citenamefont {Amitai},
  \citenamefont {Koppenh{\"{o}}fer}, \citenamefont {L{\"{o}}rch},\ and\
  \citenamefont {Bruder}}]{qad2}%
  \BibitemOpen
  \bibfield  {author} {\bibinfo {author} {\bibfnamefont {E.}~\bibnamefont
  {Amitai}}, \bibinfo {author} {\bibfnamefont {M.}~\bibnamefont
  {Koppenh{\"{o}}fer}}, \bibinfo {author} {\bibfnamefont {N.}~\bibnamefont
  {L{\"{o}}rch}},\ and\ \bibinfo {author} {\bibfnamefont {C.}~\bibnamefont
  {Bruder}},\ }\bibfield  {title} {\bibinfo {title} {Quantum effects in
  amplitude death of coupled anharmonic self-oscillators},\ }\href@noop {}
  {\bibfield  {journal} {\bibinfo  {journal} {Phys. Rev. E}\ }\textbf {\bibinfo
  {volume} {97}},\ \bibinfo {pages} {052203} (\bibinfo {year}
  {2018})}\BibitemShut {NoStop}%
\bibitem [{\citenamefont {Bastidas}\ \emph {et~al.}(2015)\citenamefont
  {Bastidas}, \citenamefont {Omelchenko}, \citenamefont {Zakharova},
  \citenamefont {Schöll},\ and\ \citenamefont {Brandes}}]{qchm}%
  \BibitemOpen
  \bibfield  {author} {\bibinfo {author} {\bibfnamefont {V.~M.}\ \bibnamefont
  {Bastidas}}, \bibinfo {author} {\bibfnamefont {I.}~\bibnamefont
  {Omelchenko}}, \bibinfo {author} {\bibfnamefont {A.}~\bibnamefont
  {Zakharova}}, \bibinfo {author} {\bibfnamefont {E.}~\bibnamefont {Schöll}},\
  and\ \bibinfo {author} {\bibfnamefont {T.}~\bibnamefont {Brandes}},\
  }\bibfield  {title} {\bibinfo {title} {Quantum signatures of chimera
  states},\ }\href@noop {} {\bibfield  {journal} {\bibinfo  {journal} {Phys.
  Rev. E}\ }\textbf {\bibinfo {volume} {92}},\ \bibinfo {pages} {062924}
  (\bibinfo {year} {2015})}\BibitemShut {NoStop}%
\bibitem [{\citenamefont {Bandyopadhyay}\ \emph {et~al.}(2020)\citenamefont
  {Bandyopadhyay}, \citenamefont {Khatun}, \citenamefont {Biswas},\ and\
  \citenamefont {Banerjee}}]{qmod}%
  \BibitemOpen
  \bibfield  {author} {\bibinfo {author} {\bibfnamefont {B.}~\bibnamefont
  {Bandyopadhyay}}, \bibinfo {author} {\bibfnamefont {T.}~\bibnamefont
  {Khatun}}, \bibinfo {author} {\bibfnamefont {D.}~\bibnamefont {Biswas}},\
  and\ \bibinfo {author} {\bibfnamefont {T.}~\bibnamefont {Banerjee}},\
  }\bibfield  {title} {\bibinfo {title} {Quantum manifestations of homogeneous
  and inhomogeneous oscillation suppression states},\ }\href@noop {} {\bibfield
   {journal} {\bibinfo  {journal} {Phys. Rev. E}\ }\textbf {\bibinfo {volume}
  {102}},\ \bibinfo {pages} {062205} (\bibinfo {year} {2020})}\BibitemShut
  {NoStop}%
\bibitem [{\citenamefont {Bandyopadhyay}\ and\ \citenamefont
  {Banerjee}(2021)}]{qrev}%
  \BibitemOpen
  \bibfield  {author} {\bibinfo {author} {\bibfnamefont {B.}~\bibnamefont
  {Bandyopadhyay}}\ and\ \bibinfo {author} {\bibfnamefont {T.}~\bibnamefont
  {Banerjee}},\ }\bibfield  {title} {\bibinfo {title} {Revival of oscillation
  and symmetry breaking in coupled quantum oscillators},\ }\href@noop {}
  {\bibfield  {journal} {\bibinfo  {journal} {Chaos}\ }\textbf {\bibinfo
  {volume} {31}},\ \bibinfo {pages} {063109} (\bibinfo {year}
  {2021})}\BibitemShut {NoStop}%
\bibitem [{\citenamefont {Bandyopadhyay}\ \emph {et~al.}(2021)\citenamefont
  {Bandyopadhyay}, \citenamefont {Khatun},\ and\ \citenamefont
  {Banerjee}}]{qturing}%
  \BibitemOpen
  \bibfield  {author} {\bibinfo {author} {\bibfnamefont {B.}~\bibnamefont
  {Bandyopadhyay}}, \bibinfo {author} {\bibfnamefont {T.}~\bibnamefont
  {Khatun}},\ and\ \bibinfo {author} {\bibfnamefont {T.}~\bibnamefont
  {Banerjee}},\ }\bibfield  {title} {\bibinfo {title} {Quantum {T}uring
  bifurcation: Transition from quantum amplitude death to quantum oscillation
  death},\ }\href@noop {} {\bibfield  {journal} {\bibinfo  {journal} {Phys.
  Rev. E}\ }\textbf {\bibinfo {volume} {104}},\ \bibinfo {pages} {024214}
  (\bibinfo {year} {2021})}\BibitemShut {NoStop}%
\bibitem [{\citenamefont {Bandyopadhyay}\ and\ \citenamefont
  {Banerjee}(2022)}]{qkerr}%
  \BibitemOpen
  \bibfield  {author} {\bibinfo {author} {\bibfnamefont {B.}~\bibnamefont
  {Bandyopadhyay}}\ and\ \bibinfo {author} {\bibfnamefont {T.}~\bibnamefont
  {Banerjee}},\ }\bibfield  {title} {\bibinfo {title} {Kerr nonlinearity
  hinders symmetry-breaking states of coupled quantum oscillators},\
  }\href@noop {} {\bibfield  {journal} {\bibinfo  {journal} {Phys. Rev. E}\
  }\textbf {\bibinfo {volume} {106}},\ \bibinfo {pages} {024216} (\bibinfo
  {year} {2022})}\BibitemShut {NoStop}%
\bibitem [{\citenamefont {Kato}\ and\ \citenamefont {Nakao}(2022)}]{kato}%
  \BibitemOpen
  \bibfield  {author} {\bibinfo {author} {\bibfnamefont {Y.}~\bibnamefont
  {Kato}}\ and\ \bibinfo {author} {\bibfnamefont {H.}~\bibnamefont {Nakao}},\
  }\bibfield  {title} {\bibinfo {title} {Turing instability in quantum
  activator-inhibitor systems},\ }\href@noop {} {\bibfield  {journal} {\bibinfo
   {journal} {Sci. Rep.}\ }\textbf {\bibinfo {volume} {12}},\ \bibinfo {pages}
  {15573} (\bibinfo {year} {2022})}\BibitemShut {NoStop}%
\bibitem [{\citenamefont {Yuge}\ \emph {et~al.}(2014)\citenamefont {Yuge},
  \citenamefont {Kamide}, \citenamefont {Yamaguchi},\ and\ \citenamefont
  {Ogawa}}]{loss-cavity}%
  \BibitemOpen
  \bibfield  {author} {\bibinfo {author} {\bibfnamefont {T.}~\bibnamefont
  {Yuge}}, \bibinfo {author} {\bibfnamefont {K.}~\bibnamefont {Kamide}},
  \bibinfo {author} {\bibfnamefont {M.}~\bibnamefont {Yamaguchi}},\ and\
  \bibinfo {author} {\bibfnamefont {T.}~\bibnamefont {Ogawa}},\ }\bibfield
  {title} {\bibinfo {title} {Cavity-loss induced plateau in coupled cavity qed
  array},\ }\href@noop {} {\bibfield  {journal} {\bibinfo  {journal} {Journal
  of the Physical Society of Japan}\ }\textbf {\bibinfo {volume} {83}},\
  \bibinfo {pages} {123001} (\bibinfo {year} {2014})}\BibitemShut {NoStop}%
\bibitem [{\citenamefont {Aspelmeyer}\ \emph {et~al.}(2014)\citenamefont
  {Aspelmeyer}, \citenamefont {Kippenberg},\ and\ \citenamefont
  {Marquardt}}]{opto-revmodphys}%
  \BibitemOpen
  \bibfield  {author} {\bibinfo {author} {\bibfnamefont {M.}~\bibnamefont
  {Aspelmeyer}}, \bibinfo {author} {\bibfnamefont {T.~J.}\ \bibnamefont
  {Kippenberg}},\ and\ \bibinfo {author} {\bibfnamefont {F.}~\bibnamefont
  {Marquardt}},\ }\bibfield  {title} {\bibinfo {title} {Cavity optomechanics},\
  }\href@noop {} {\bibfield  {journal} {\bibinfo  {journal} {Rev. Mod. Phys.}\
  }\textbf {\bibinfo {volume} {86}},\ \bibinfo {pages} {1391} (\bibinfo {year}
  {2014})}\BibitemShut {NoStop}%
\bibitem [{\citenamefont {Fitzgerald}\ \emph {et~al.}(2021)\citenamefont
  {Fitzgerald}, \citenamefont {Manjeshwar}, \citenamefont {Wieczorek},\ and\
  \citenamefont {Tassin}}]{opto-loss}%
  \BibitemOpen
  \bibfield  {author} {\bibinfo {author} {\bibfnamefont {J.~M.}\ \bibnamefont
  {Fitzgerald}}, \bibinfo {author} {\bibfnamefont {S.~K.}\ \bibnamefont
  {Manjeshwar}}, \bibinfo {author} {\bibfnamefont {W.}~\bibnamefont
  {Wieczorek}},\ and\ \bibinfo {author} {\bibfnamefont {P.}~\bibnamefont
  {Tassin}},\ }\bibfield  {title} {\bibinfo {title} {Cavity optomechanics with
  photonic bound states in the continuum},\ }\href@noop {} {\bibfield
  {journal} {\bibinfo  {journal} {Phys. Rev. Research}\ }\textbf {\bibinfo
  {volume} {3}},\ \bibinfo {pages} {013131} (\bibinfo {year}
  {2021})}\BibitemShut {NoStop}%
\bibitem [{\citenamefont {Pedersen}\ \emph {et~al.}(2002)\citenamefont
  {Pedersen}, \citenamefont {Strasser}, \citenamefont {Heber}, \citenamefont
  {Rappaport},\ and\ \citenamefont {Zajfman}}]{loss-ion}%
  \BibitemOpen
  \bibfield  {author} {\bibinfo {author} {\bibfnamefont {H.~B.}\ \bibnamefont
  {Pedersen}}, \bibinfo {author} {\bibfnamefont {D.}~\bibnamefont {Strasser}},
  \bibinfo {author} {\bibfnamefont {O.}~\bibnamefont {Heber}}, \bibinfo
  {author} {\bibfnamefont {M.~L.}\ \bibnamefont {Rappaport}},\ and\ \bibinfo
  {author} {\bibfnamefont {D.}~\bibnamefont {Zajfman}},\ }\bibfield  {title}
  {\bibinfo {title} {Stability and loss in an ion-trap resonator},\ }\href@noop
  {} {\bibfield  {journal} {\bibinfo  {journal} {Phys. Rev. A}\ }\textbf
  {\bibinfo {volume} {65}},\ \bibinfo {pages} {042703} (\bibinfo {year}
  {2002})}\BibitemShut {NoStop}%
\bibitem [{\citenamefont {Blais}\ \emph {et~al.}(2021)\citenamefont {Blais},
  \citenamefont {Grimsmo}, \citenamefont {Girvin},\ and\ \citenamefont
  {Wallraff}}]{cQED}%
  \BibitemOpen
  \bibfield  {author} {\bibinfo {author} {\bibfnamefont {A.}~\bibnamefont
  {Blais}}, \bibinfo {author} {\bibfnamefont {A.~L.}\ \bibnamefont {Grimsmo}},
  \bibinfo {author} {\bibfnamefont {S.~M.}\ \bibnamefont {Girvin}},\ and\
  \bibinfo {author} {\bibfnamefont {A.}~\bibnamefont {Wallraff}},\ }\bibfield
  {title} {\bibinfo {title} {Circuit quantum electrodynamics},\ }\href@noop {}
  {\bibfield  {journal} {\bibinfo  {journal} {Rev. Mod. Phys.}\ }\textbf
  {\bibinfo {volume} {93}},\ \bibinfo {pages} {025005} (\bibinfo {year}
  {2021})}\BibitemShut {NoStop}%
\bibitem [{\citenamefont {Carmichael}(1999)}]{carmichael}%
  \BibitemOpen
  \bibfield  {author} {\bibinfo {author} {\bibfnamefont {H.~J.}\ \bibnamefont
  {Carmichael}},\ }\href@noop {} {\emph {\bibinfo {title} {Statistical Methods
  in Quantum Optics 1}}}\ (\bibinfo  {publisher} {Springer},\ \bibinfo {year}
  {1999})\BibitemShut {NoStop}%
\bibitem [{\citenamefont {Johansson}\ \emph {et~al.}(2012)\citenamefont
  {Johansson}, \citenamefont {Nation},\ and\ \citenamefont {Nori}}]{qutip1}%
  \BibitemOpen
  \bibfield  {author} {\bibinfo {author} {\bibfnamefont {J.}~\bibnamefont
  {Johansson}}, \bibinfo {author} {\bibfnamefont {P.}~\bibnamefont {Nation}},\
  and\ \bibinfo {author} {\bibfnamefont {F.}~\bibnamefont {Nori}},\ }\bibfield
  {title} {\bibinfo {title} {Qutip: An open-source python framework for the
  dynamics of open quantum systems},\ }\href@noop {} {\bibfield  {journal}
  {\bibinfo  {journal} {Comp. Phys. Comm.}\ }\textbf {\bibinfo {volume}
  {183}},\ \bibinfo {pages} {1760} (\bibinfo {year} {2012})}\BibitemShut
  {NoStop}%
\bibitem [{\citenamefont {Johansson}\ \emph {et~al.}(2013)\citenamefont
  {Johansson}, \citenamefont {Nation},\ and\ \citenamefont {Nori}}]{qutip}%
  \BibitemOpen
  \bibfield  {author} {\bibinfo {author} {\bibfnamefont {J.}~\bibnamefont
  {Johansson}}, \bibinfo {author} {\bibfnamefont {P.}~\bibnamefont {Nation}},\
  and\ \bibinfo {author} {\bibfnamefont {F.}~\bibnamefont {Nori}},\ }\bibfield
  {title} {\bibinfo {title} {Qutip 2: A python framework for the dynamics of
  open quantum systems},\ }\href@noop {} {\bibfield  {journal} {\bibinfo
  {journal} {Comp. Phys. Comm.}\ }\textbf {\bibinfo {volume} {184}},\ \bibinfo
  {pages} {1234} (\bibinfo {year} {2013})}\BibitemShut {NoStop}%
\bibitem [{\citenamefont {Borrielli}\ \emph {et~al.}(2015)\citenamefont
  {Borrielli}, \citenamefont {Pontin}, \citenamefont {Cataliotti},
  \citenamefont {Marconi}, \citenamefont {Marin}, \citenamefont {Marino},
  \citenamefont {Pandraud}, \citenamefont {Prodi}, \citenamefont {Serra},\ and\
  \citenamefont {Bonaldi}}]{low-loss}%
  \BibitemOpen
  \bibfield  {author} {\bibinfo {author} {\bibfnamefont {A.}~\bibnamefont
  {Borrielli}}, \bibinfo {author} {\bibfnamefont {A.}~\bibnamefont {Pontin}},
  \bibinfo {author} {\bibfnamefont {F.~S.}\ \bibnamefont {Cataliotti}},
  \bibinfo {author} {\bibfnamefont {L.}~\bibnamefont {Marconi}}, \bibinfo
  {author} {\bibfnamefont {F.}~\bibnamefont {Marin}}, \bibinfo {author}
  {\bibfnamefont {F.}~\bibnamefont {Marino}}, \bibinfo {author} {\bibfnamefont
  {G.}~\bibnamefont {Pandraud}}, \bibinfo {author} {\bibfnamefont {G.~A.}\
  \bibnamefont {Prodi}}, \bibinfo {author} {\bibfnamefont {E.}~\bibnamefont
  {Serra}},\ and\ \bibinfo {author} {\bibfnamefont {M.}~\bibnamefont
  {Bonaldi}},\ }\bibfield  {title} {\bibinfo {title} {Low-loss optomechanical
  oscillator for quantum-optics experiments},\ }\href@noop {} {\bibfield
  {journal} {\bibinfo  {journal} {Phys. Rev. Applied}\ }\textbf {\bibinfo
  {volume} {3}},\ \bibinfo {pages} {054009} (\bibinfo {year}
  {2015})}\BibitemShut {NoStop}%
\bibitem [{\citenamefont {{Borrielli}}\ \emph {et~al.}(2015)\citenamefont
  {{Borrielli}}, \citenamefont {{Pontin}}, \citenamefont {{Cataliotti}},
  \citenamefont {{Marconi}}, \citenamefont {{Marin}}, \citenamefont {{Marino}},
  \citenamefont {{Pandraud}}, \citenamefont {{Prodi}}, \citenamefont
  {{Serra}},\ and\ \citenamefont {{Bonaldi}}}]{low-loss2}%
  \BibitemOpen
  \bibfield  {author} {\bibinfo {author} {\bibfnamefont {A.}~\bibnamefont
  {{Borrielli}}}, \bibinfo {author} {\bibfnamefont {A.}~\bibnamefont
  {{Pontin}}}, \bibinfo {author} {\bibfnamefont {F.~S.}\ \bibnamefont
  {{Cataliotti}}}, \bibinfo {author} {\bibfnamefont {L.}~\bibnamefont
  {{Marconi}}}, \bibinfo {author} {\bibfnamefont {F.}~\bibnamefont {{Marin}}},
  \bibinfo {author} {\bibfnamefont {F.}~\bibnamefont {{Marino}}}, \bibinfo
  {author} {\bibfnamefont {G.}~\bibnamefont {{Pandraud}}}, \bibinfo {author}
  {\bibfnamefont {G.~A.}\ \bibnamefont {{Prodi}}}, \bibinfo {author}
  {\bibfnamefont {E.}~\bibnamefont {{Serra}}},\ and\ \bibinfo {author}
  {\bibfnamefont {M.}~\bibnamefont {{Bonaldi}}},\ }\bibfield  {title} {\bibinfo
  {title} {{Low loss optomechanical cavities based on silicon oscillator}},\
  }in\ \href@noop {} {\emph {\bibinfo {booktitle} {Smart Sensors, Actuators,
  and MEMS VII; and Cyber Physical Systems}}},\ \bibinfo {series} {Society of
  Photo-Optical Instrumentation Engineers (SPIE) Conference Series}, Vol.\
  \bibinfo {volume} {9517},\ \bibinfo {editor} {edited by\ \bibinfo {editor}
  {\bibfnamefont {J.~L.}\ \bibnamefont {{S{\'a}nchez-Rojas}}}\ and\ \bibinfo
  {editor} {\bibfnamefont {R.}~\bibnamefont {{Brama}}}}\ (\bibinfo {year}
  {2015})\ p.\ \bibinfo {pages} {95171O}\BibitemShut {NoStop}%
\end{thebibliography}
%apsrev4-2.bst 2019-01-14 (MD) hand-edited version of apsrev4-1.bst
%Control: key (0)
%Control: author (8) initials jnrlst
%Control: editor formatted (1) identically to author
%Control: production of article title (0) allowed
%Control: page (0) single
%Control: year (1) truncated
%Control: production of eprint (0) enabled
\providecommand{\noopsort}[1]{}\providecommand{\singleletter}[1]{#1}%
\end{document}